\documentclass{PoS}
\usepackage{url}

\title{
Many flavor QCD as exploration of the walking behavior with the approximate IR fixed point}

\ShortTitle{Exploration of the walking behavior with the approximate IR fixed point}

\author{
Yasumichi Aoki$^a$, \, Tatsumi Aoyama$^a$, \, Masafumi Kurachi$^a$, \, Toshihide Maskawa$^a$, \, \speaker{Kei-ichi Nagai}\thanks{E-mail: keiichi.nagai@kmi.nagoya-u.ac.jp} \,$^a$, \, Hiroshi Ohki$^a$, \, Akihiro Shibata$^b$, \, Koichi Yamawaki$^a$  and \, Takeshi Yamazaki$^a$ 

\hspace{55mm} LatKMI Collaboration\\
        $^a$Kobayashi-Maskawa Institute for the Origin of Particles and the Universe (KMI), 
        Nagoya University, Nagoya, 464-8602, Japan\\
        $^b$Computing Research Center, High Energy Accelerator Research Organization (KEK), 
        Tsukuba, 305-0801, Japan \\
}

\abstract{
We present the first report of the LatKMI collaboration 
on the the lattice QCD simulation performed at the KMI computer,  ``$\varphi$'',    
for the cases of 4 flavors and 8 flavors, 
the latter being expected to be a candidate for 
the walking technicolor having 
an approximate scale invariance near the infrared fixed point. 
The simulation was carried out based on 
the highly improved staggered quark (HISQ) action. 
In this proceedings, 
we report preliminary results on 
the spectrum, 
analyzed through the chiral perturbation theory 
and the finite-size hyperscaling.
We observe qualitatively different behavior of  the 8-flavor case 
in contrast to the 4-flavor case 
which shows clear indication of  the hadronic phase 
as in the usual QCD.
}

\FullConference{ The XXIX International Symposium on Lattice Field Theory - Lattice 2011\\
July 10-16, 2011\\
Squaw Valley, Lake Tahoe, California}

\begin{document}

\section{Introduction}
\label{sec:intro}

The origin of mass is the most urgent issue of the particle physics today.
One of the candidates for the theory 
beyond the Standard Model towards 
that problem is the walking technicolor 
which is the strongly coupled gauge theory 
having a large anomalous dimension 
$\gamma_m \simeq 1$  
and approximate scale invariance 
due to the almost  non-running (walking) coupling~\cite{Yamawaki:1985zg,Akiba:1985rr}.
The walking behavior is in fact  realized in the QCD 
with  large number of (massless) flavors  $N_f$ 
which possesses 
Caswell-Banks-Zaks infrared fixed point (IRFP)~\cite{beta}
for  the value larger than $N_ f \simeq 8.0$ 
in the two-loop beta function. 
The exact IRFP would be  washed out by  the dynamical generation of a quark mass $m$ 
in the very infrared region $\mu<m$
for $N_f < N_f^{cr}$, 
$N_f^{cr}$ being the critical number. 
However,  for  $N_f$ very close to $N_f^{cr}$,  $m$ 
could be much smaller than the intrinsic scale  $\Lambda$  ($\gg  m $), 
an analogue of $\Lambda_{\rm QCD}$, 
beyond which the coupling runs as the asymptotically free theory,  
 so that the coupling remains almost non-running 
 for the wide infrared region $m<\mu<\Lambda$ 
 as a remnant of the would-be IRFP. 
 The case $N>N_f^{cr}$ is called conformal window, 
 although conformality is broken 
 in the ultraviolet  asymptotically free region beyond $\Lambda$.
 The critical number $N_f^{cr}$ was estimated 
 as $N_f^{cr} \simeq 11.9$~\cite{Appelquist:1996dq} by comparing the two-loop IRFP value 
 with the critical coupling of  
 the ladder Schwinger-Dyson equation analysis~\cite{Maskawa:1974vs}.

Although the above results from the two-loop 
and ladder approximation are very suggestive,  
the relevant dynamics is obviously of  non-perturbative nature,
we would need fully non-perturbative studies. 
Among others the lattice simulations developed in the lattice QCD 
would be the most powerful tool 
for that purpose.
Our group, LatKMI Collaboration, 
was organized for such studies on walking technicolor 
as a candidate for the theory beyond the Standard Model.
The immediate issues are:
What is the critical number $N_f^{cr}$?
What is the signatures of the  walking theory expected to be slightly smaller than $N_f^{cr}$?
The above two-loop and ladder studies suggest 
that the walking theory if existed would be in between $N_f=8$ and $N_f=12$.
The $N_f=8$ in particular is interesting 
from the model-building point of view: 
The typical technicolor model~\cite{Farhi:1980xs}  is the so-called 
one-family model (Farhi-Susskind model)  
which has a one-family of the colored techni-fermions (techni-quarks) 
and the uncolored one (techni-leptons)
corresponding to the each family of the SM quarks and leptons. 
It can embed the technicolor gauge 
and the gauged three generations of the SM fermions 
into a single gauge group (Extended Technicolor) 
and thus is the most straightforward way to accommodate the techni-fermions 
and the SM fermions into a simple scheme.  
Thus if the $N_f=8$ turns out to be a  walking theory, 
it would be a great message for the phenomenology 
to be tested by the on-going LHC. 

To date,
some groups~\cite{appelquist,iwasaki,Brown:1992fz,Fodor:2009wk,Deuzeman:2008sc,Jin:2010vm} 
carried out lattice studies on 8 flavors;
Ref.~\cite{appelquist} computed the running coupling constant in $N_f=8$
by the Schr\"odinger functional method in staggered fermion case
and reported that $N_f=8$ is in the chiral broken phase.
Refs.~\cite{iwasaki,Brown:1992fz,Fodor:2009wk,Deuzeman:2008sc,Jin:2010vm} 
investigated the hadron spectrum with 
the standard Wilson fermions~\cite{iwasaki},
the Stout improved staggered fermions~\cite{Fodor:2009wk},
the Asqtad improved staggered fermions~\cite{Deuzeman:2008sc}
and the na\"ive staggered fermions~\cite{Brown:1992fz,Jin:2010vm}.
The Ref.~\cite{iwasaki} concluded the $N_f=8$ is in the conformal window,
but other groups concluded
that the $N_f=8$ resides on the chiral broken phase.
Even if $N_f=8$ is in the chiral broken phase,
it is not clear whether the behavior of this system is {\it QCD-like} 
or the walking with the large anomalous mass dimension.
Nobody has investigated the possibility that $N_f=8$ is in the walking.

We  simulate 8-flavor QCD with alternative lattice fermion, HISQ,
in which the flavor symmetry in the staggered fermion is improved
and, of course, it is expected that the behavior towards the continuum limit is improved.
We show the preliminary result of the hadron spectrum
and analyzed the data based on  the hyperscaling~\cite{DelDebbio:2010ze} 
as well as the chiral perturbation theory (ChPT).
From the hyperscaling analysis,
we derive the anomalous mass dimension $\gamma_m$.
We observe  qualitatively different behavior of  the 8-flavor case 
in contrast to the 4-flavor case which shows
clear indication of  the hadronic phase as in the usual QCD.

\section{Simulation}
\label{sec:simu}

In our simulation,
we use the tree level Symanzik gauge action
and the highly improved staggered quark (HISQ) action
without the tadpole improvement 
and the mass correction in the Naik term.
See Ref.~\cite{Follana:2006rc} for the detail of the HISQ action.
We use the MILC code~\cite{milc}
with modifications 
to simulate $N_f=4 n$ HISQ 
by using the standard Hybrid Monte-Carlo (HMC) algorithm.
We computed the hadron spectrum
as the global survey in the parameter region
and 
we obtained $M_\pi$, $M_\rho$, $f_\pi$	
and $\langle \bar{\psi}\psi \rangle$
as the basic observable.

The simulation for $N_f=4$ is carried out 
at $\beta(=6/g^2)$=3.5, 3.6, 3.7 and 3.8
for various quark masses
on $12^3 \times 16$ and $16^3 \times 24$.
We took over 1000 trajectories on the small lattice
and about 600 trajectories on the large lattice
in $N_f=4$ case.
The simulation for $N_f=8$ is carried out 
at $\beta(=6/g^2)$=3.6, 3.7 and 3.8
for various quark masses
on $12^3 \times 32$ and $24^3 \times 32$,
and for  $m_f=$0.02 and 0.04 on $30^3 \times 40$ at $\beta=3.7$.
We took about 1000 trajectories on $12^3 \times 32$, 
between 200 and 400 trajectories on $24^3 \times 32$
and 600 trajectories on $30^3 \times 40$.

See Ref.~\cite{latkmi1216} for our simulation in $N_f=12$ and 16.

\section{Spectrum}
\label{sec:spectrum}

In this section,
we show preliminary results
of  ChPT analysis
and the finite-size hyperscaling analysis
in 4- and 8-flavor cases.

\subsection{$N_f=4$}
\label{sec:nf4}

\begin{figure}[htb]
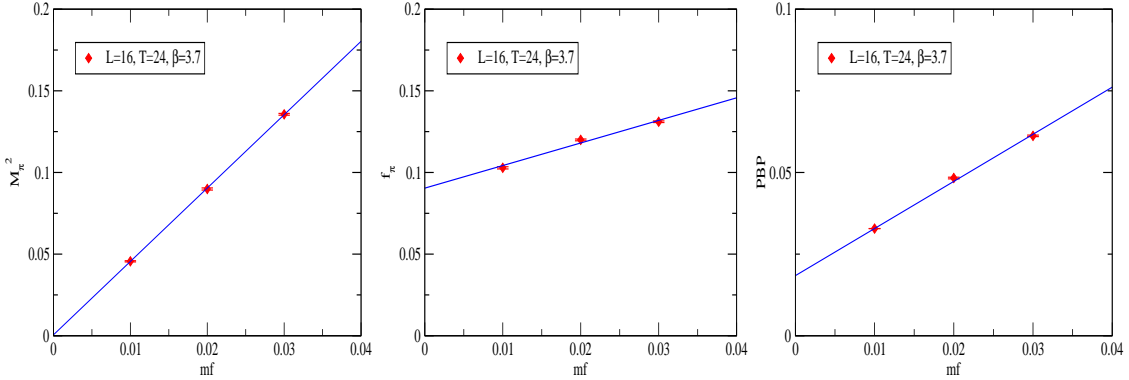
 
\center{
\includegraphics[width=4.9cm,height=5.0cm,trim=0 0 0 0,clip=true]{plots/Nf04L16T24B3.7mpi2.eps}
\includegraphics[width=4.9cm,height=5.0cm,trim=0 0 0 0,clip=true]{plots/Nf04L16T24B3.7fpi.eps}
\includegraphics[width=4.9cm,height=5.0cm,trim=0 0 0 0,clip=true]{plots/Nf04L16T24B3.7PBP.eps}
}
\caption{
In $N_f=4$ $SU(3)$ gauge theory on $16^3 \times 24$ at $\beta=3.7$; 
{\bf Left:} $M_\pi^2$ as functions of $m_f$,
{\bf Center:} $f_\pi$ as functions of $m_f$,
{\bf Right:} $\langle \bar{\psi}\psi \rangle$ as functions of $m_f$.
In all panels, the blue-solid line is the linear fit.
}
\label{fig:nf4}
\end{figure}

The result of  $N_f=4$ is shown in Fig.~\ref{fig:nf4}.
The data on $16^3\times 24$ at $\beta=3.7$,
the pion mass squared, the decay constant and the chiral condensate 
are plotted on the panel from the left to the right respectively.
$M_\pi^2$ is proportional to  $m_f$.
$f_\pi$ and $\langle \bar{\psi}\psi \rangle$ 
have the non zero value 
in the chiral limit.
Thus,  the $N_f=4$ QCD has the property of  the chiral broken phase
and this is regarded as the signal of the chiral broken phase
in the dynamical case of lattice QCD.

\subsection{$N_f=8$}
\label{subsec:nf8}

\begin{figure}[tbh]
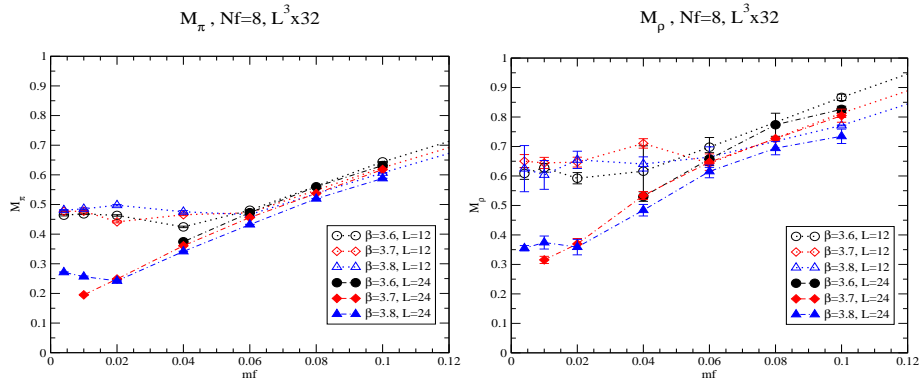
 
\center{
\includegraphics[width=6.0cm,height=5.0cm,trim=0 0 0 0,clip=true]{plots/mpimf_allL.eps}
\includegraphics[width=6.0cm,height=5.0cm,trim=0 0 0 0,clip=true]{plots/mrhomf_allL.eps}
}
\caption{
In $N_f=8$ $SU(3)$ gauge theory
on $12^3 \times 32$ and $24^3 \times 32$
at $\beta=3.6$, 3.7 and 3.8; 
{\bf Left panel:} $M_\pi$ as functions of $m_f$, 
{\bf Right panel:} $M_\rho$ as functions of $m_f$. 
}
\label{fig:spectrum}
\end{figure}

In this subsection,
we analyze $N_f=8$ system
by  ChPT and the finite-size hyperscaling relation.
In Fig.~\ref{fig:spectrum}, 
as the typical example of the raw data,
we plotted the preliminary data 
of $M_\pi$ and $M_\rho$ as a function of the quark mass $m_f$
on  $12^3 \times 32$ and $24^3 \times 32$.
It is shown that $M_\pi$ and $M_\rho$ are similar behavior,
that is, the plateau appears  in $m_f \lesssim 0.06$
on small lattice ($12^3 \times 32$) at all $\beta$s
and  in $m_f \lesssim 0.02$ on large lattice ($24^3 \times 32$) at $\beta=3.8$.
In the following,
we attempt ChPT and the finite-size hyperscaling analysis
by using these data.

\subsubsection{ChPT analysis}
\label{subsec:chpt}

\begin{figure}[htb]
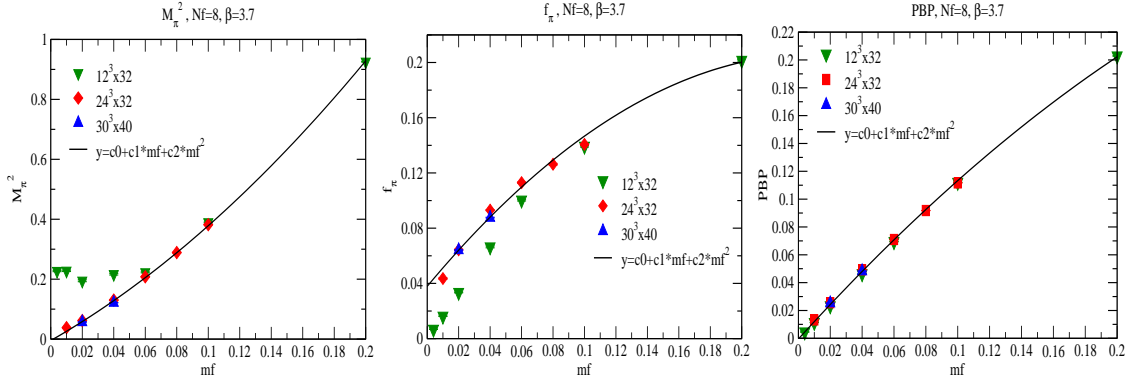
 
\center{
\includegraphics[width=4.9cm,height=5.0cm,trim=0 0 0 0,clip=true]{plots/mpi2fit_b37.eps}
\includegraphics[width=4.9cm,height=5.0cm,trim=0 0 0 0,clip=true]{plots/fpifit_b37.eps}
\includegraphics[width=4.9cm,height=5.0cm,trim=0 0 0 0,clip=true]{plots/pbpfit_b37.eps}
}
\caption{
In $SU(3)$ gauge theory with 8-flavors
on $12^3 \times 32$, $24^3 \times 32$ 
and $30^3 \times 40$ at $\beta=3.7$; 
{\bf Left:} $M_\pi^2$ as functions of $m_f$. 
$\chi^2/d.o.f.=42$.
{\bf Center:} $f_\pi$. $\chi^2/d.o.f.=105$.
{\bf Right:} $\langle \bar{\psi}\psi \rangle$.  $\chi^2/d.o.f.=154$; \,
In all panels, the solid line stands for the quadratic-fit.
}
\label{fig:fit-b3.7}
\end{figure}

We analyze the $N_f=8$ data by ChPT.
Here, we pick up the data at $\beta=3.7$,
which has the $30^3 \times 40$ volume data.
Fig.~\ref{fig:fit-b3.7} shows 
$M_\pi^2$, $f_\pi$ and $\langle \bar{\psi}\psi \rangle$.
For each  $m_f$,
the largest volume data is used for the quadratic-fit corresponding to  ChPT analysis.
The data corresponding to the plateau is not included into the fit.

As the fit result, 
$M_\pi^2$ in the chiral limit is consistent with zero.
This is fairly independent of the fit range.
The quadratic-fit of $f_\pi$ in the chiral limit indicates non-zero value.
These properties are consistent with the chiral broken phase.
On the other hand,
the condensate is not inconsistent with zero in the chiral limit, 
in contrast to $N_f=4$ case (compare Fig.~\ref{fig:fit-b3.7} with Fig.~\ref{fig:nf4}).

The definite conclusion cannot be given here
because $\chi^2/d.o.f. \sim O(10)$  in ChPT fit 
and because it is difficult to take the infinite volume limit.
Even if $N_f=8$ is in the broken phase,
there is a possibility that  the walking behavior
as a remnant of the conformality
can be observed in the form of hyperscaling relation.

\subsubsection{Finite-size hyperscaling analysis}
\label{subsec:hyper}

If the system is in the conformal window,
physical quantities, $M_H$,  are described 
by the finite-size hyperscaling relation~\cite{DelDebbio:2010ze} ; 
$L M_H = {\cal F}(X)$  where ${\cal F}(X)$ is unknown function 
with the scaling variable $X=L m_f^{\frac{1}{1+\gamma}}$.
The $\gamma$ in this equation is defined as the anomalous mass dimension.
We carry out the hyperscaling analysis with our data of $M_H=\{M_\pi, f_\pi, M_\rho\}$  
by the following fit function
\begin{equation}
L M_H =  c_0 + c_1 L m_f^{\frac{1}{1+\gamma}}  \, .
\label{eq:hs_linear}
\end{equation}

\begin{figure}[tbh]
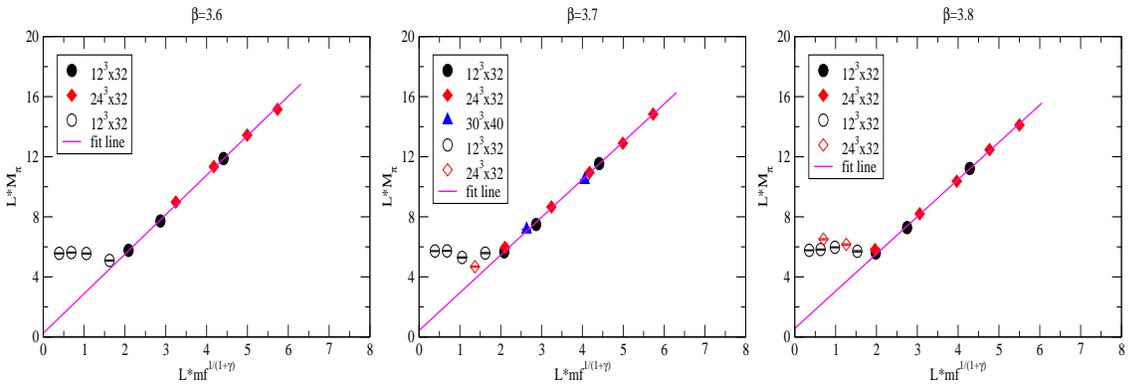
 
\center{
\includegraphics[width=4.9cm,height=5.0cm,trim=0 0 0 0,clip=true]{plots/mpiL-b3.6.eps}
\includegraphics[width=4.9cm,height=5.0cm,trim=0 0 0 0,clip=true]{plots/mpiL-b3.7.eps}
\includegraphics[width=4.9cm,height=5.0cm,trim=0 0 0 0,clip=true]{plots/mpiL-b3.8.eps}
}
\caption{
{\bf Left:} The hyperscaling relation of $L M_\pi$ as functions of $L m_f^{1/1+\gamma}$ in $N_f=8$
at $\beta=3.6$,  $\chi^2/d.o.f.=41.7$.
{\bf Center:} at $\beta=3.7$, $\chi^2/d.o.f.=39.8$.
{\bf Right:} at $\beta=3.8$,  $\chi^2/d.o.f.=37.3$. \,
At all $\beta$, $\gamma(M_\pi) \sim 0.6$. \, 
The filled symbol is applied to the hyperscaling fit.
}
\label{fig:nf8hs_mpi}
\end{figure}

\begin{figure}[tbh]
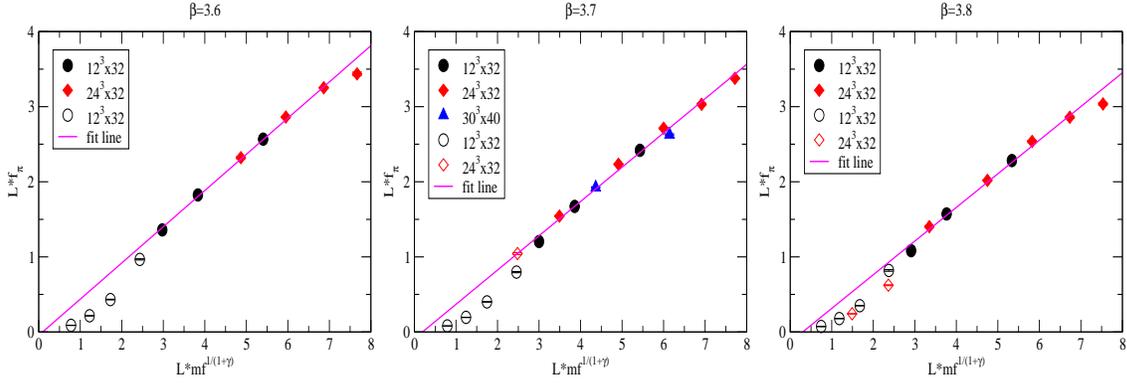
 
\center{
\includegraphics[width=4.9cm,height=5.0cm,trim=0 0 0 0,clip=true]{plots/fpiL-b3.6.eps}
\includegraphics[width=4.9cm,height=5.0cm,trim=0 0 0 0,clip=true]{plots/fpiL-b3.7.eps}
\includegraphics[width=4.9cm,height=5.0cm,trim=0 0 0 0,clip=true]{plots/fpiL-b3.8.eps}
}
\caption{
{\bf Left:} The hyperscaling relation of $L f_\pi$ as functions of $L m_f^{1/1+\gamma}$ in $N_f=8$
at $\beta=3.6$, $\chi^2/d.o.f.=30.9$.
{\bf Center:} at $\beta=3.7$, $\chi^2/d.o.f.=57.8$.
{\bf Right:} at $\beta=3.8$, $\chi^2/d.o.f.=55.1$.\,
At all $\beta$, $\gamma(f_\pi) \sim 1.0$.
}
\label{fig:nf8hs_fpi}
\end{figure}

\begin{figure}[tbh]
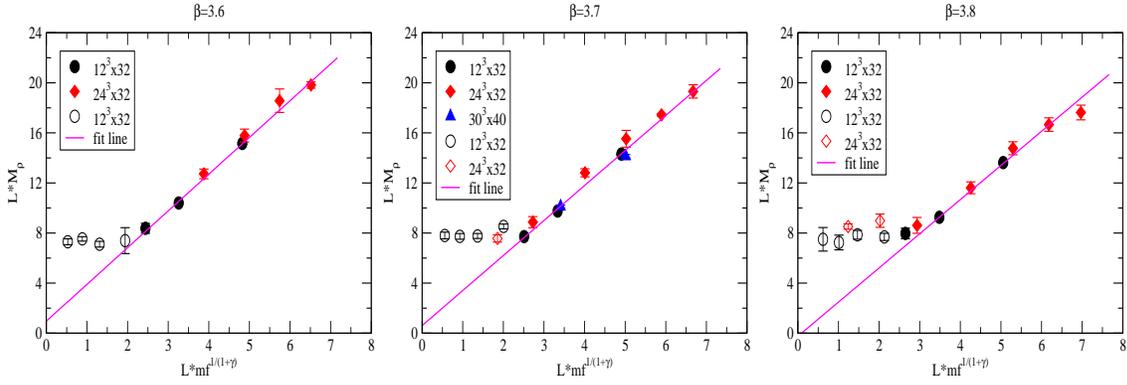
 
\center{
\includegraphics[width=4.9cm,height=5.0cm,trim=0 0 0 0,clip=true]{plots/mrhoL-b3.6.eps}
\includegraphics[width=4.9cm,height=5.0cm,trim=0 0 0 0,clip=true]{plots/mrhoL-b3.7.eps}
\includegraphics[width=4.9cm,height=5.0cm,trim=0 0 0 0,clip=true]{plots/mrhoL-b3.8.eps}
}
\caption{
{\bf Left:} The hyperscaling relation of $L M_\rho$ as functions of $L m_f^{1/1+\gamma}$  in $N_f=8$ 
at $\beta=3.6$, $\chi^2/d.o.f.=1.3$.
{\bf Center:} at $\beta=3.7$, $\chi^2/d.o.f.=5.7$.
{\bf Right:} at $\beta=3.8$, $\chi^2/d.o.f.=2.6$. \,
At all $\beta$, $\gamma(M_\rho) \sim 0.8$.
}
\label{fig:nf8hs_mrho}
\end{figure}

Figs.~\ref{fig:nf8hs_mpi}, ~\ref{fig:nf8hs_fpi} and ~\ref{fig:nf8hs_mrho}
are the finite-size hyperscaling result with this fitting 
of $L M_\pi$, $L f_\pi$ and $L M_\rho$ respectively.
The filled symbol is applied to the hyperscaling fitting.
These data on various lattice sizes
align 
at an optimal value of $\gamma$.
Although the fit quality is more or less the same level as that of the ChPT fit,
we extracted the $\gamma$-value
$\gamma(M_\pi) \sim 0.6$,
$\gamma(f_\pi) \sim 1.0$ and
$\gamma(M_\rho) \sim 0.8$
at all $\beta$, 
which are not universal.
Therefore,
our result of 8 flavors does not show the clear  signature of the conformal window.
Still, 
this situation in $N_f=8$ is qualitatively different from the case of  $N_f=4$.  
Actually we did not find 
even an alignment of the hyperscaling for each physical quantity  in $N_f=4$.

\section{Summary}
\label{sec:summary}

We have made simulations of lattice QCD
with 4 and 8 flavors 
by using the HISQ action.
We obtained the following result;
The $N_f=4$ QCD is in good agreement with the chiral broken phase.
The $N_f=8$, on the other hand,
does not seem to be inconsistent with both ChPT and the finite-size hyperscaling,
with $\chi^2/d.o.f.$ not being small for both analyses.
We extracted the $\gamma$-value from the hyperscaling analysis,
$\gamma(M_\pi) \sim 0.6$,
$\gamma(f_\pi) \sim 1.0$ and
$\gamma(M_\rho) \sim 0.8$
at all $\beta$, 
which are not universal.
Therefore,
our result of 8 flavors does not show the clear  signature of the conformal window.
This may be an indication of the walking behavior
that appears 
in the broken phase 
just below the edge of the conformal window.

We should mention that there are several possible systematic uncertainties not considered 
in this report;
Our data include those with slightly different aspect ratio,
the ratio of the temporal length to the spatial length.
Our result of the finite-size hyperscaling analysis 
may suffer from the systematic error 
due to the different aspect ratio.
Furthermore,  as pointed out in Ref.~\cite{Aoki:2012ve},
there exists the mass correction in the hyperscaling relation
for the heavy quark region.
Also the linear ansatz adapted in Eq.~(\ref{eq:hs_linear})  may not be sufficient
to fit our data.
To improve the situation for better understanding,
we will accumulate more data
for various fermion masses and $\beta$'s
on larger lattices,
and  carry out detailed analysis using those data.

\acknowledgments
Numerical simulation has been carried out on 
the supercomputer system ``$\varphi$'',
which is installed at KMI for the studies beyond the Standard Model.
We thank Katsuya Hasebe for the useful discussion.
This work is supported
by the JSPS Grant-in-Aid for Scientific Research (S) No.22224003, 
 (C) No.23540300 (K.Y.) and (C) No.21540289 (Y.A.),
and also by Grants-in-Aid of the Japanese Ministry for Scientific Research 
on Innovative Areas No. 23105708 (T.Y.).


\end{document}